# A high precision falling-ball viscometer using a fast camera

Neal Samuel Border, Aiden Reilly, Addison Miller, Shirin Jamali, and A. K. M. Newaz

Department of Physics and Astronomy, San Francisco State University, San Francisco, California 94132, USA

(June 29$^{th}$, 2020)

This paper describes a simple and inexpensive method of measuring viscosity of a Newtonian fluid using the ball drop technique and an inexpensive point and shoot ~1000 frame per second camera. We successfully measured the viscosity of glycerol and glycerol-water mixture with high precision. We used three different size copper balls of diameters 0.8 mm, 1.59 mm, and 2.38 mm to check the accuracy of the measured viscosity in different concentrations of glycerol-water mixer solutions ranging from 50% to 100% (pure glycerol). Our measurements are in excellent agreement with the measurements conducted by other standard techniques. The simple and inexpensive techniques and physics we present in this manuscript can be employed to create a simple viscosity measurement setup for learning about complex fluid mechanics even at the undergraduate laboratory and high school teaching laboratory.

**Introduction**

Viscosity is a fundamental property of a fluid and is observable in everyday experiences from pouring paints, water, cooking oils, and washing detergents to mixing chemical solutions or suspensions. Precise determination of viscosity is critically important for various industrial applications from liquid transport, filtration, lubrication, drug formulation and delivery.

Commercially available common viscometers include a capillary, vibrational, and rotational viscometer. The majority of these methods require sophisticated equipment and a data acquisition system to produce precise viscosity measurements. However, the method of a falling ball through a fluid can use common lab equipment and produce a precise measurements at a low cost. Ball drop viscometers are used widely in different applications from oils in kitchen and cars to cell biology and medicine.[1-7] The precision of a ball drop viscometer depends on the measurement of the ball drop time of flight. Commonly used time measurement devices, such as stopwatch or simple camera, do not have enough time resolutions to measure the fast motion of a ball. To obtain high time resolution, we employed a simple point and shoot camera with ~1000 frames per second and record the time of flight of a dropping ball with millisecond (ms) precision. This allowed us to design and build a ball drop viscometer for high precision measurements.

Here we measured the viscosity of glycerol and glycerol water solutions as a test fluid for the ball drop technique. We calculated the viscosity by dropping Cu balls of known densities and measuring its velocity by recording the ball motion using a high-speed camera (CASIO- EXILIM EX-ZR1700SR).

The high velocity of a fluid creates turbulent flow and low velocity of a fluid creates a laminar flow. Turbulent flow is an unsteady and non-uniform flow, whereas laminar flow is a sheet of fluid flowing with minimal disruption. Whether the fluid is considered laminar or turbulent is characterized by its Reynolds number, which is the ratio of the kinetic force to the viscous force. Since the fluid in our experiment is static, we can assume that the sphere motion is not creating any turbulent motion.

**Theoretical Background**

When a Cu sphere is falling through a fluid there is an opposing force known as the drag force ($f$). According to Stoke's law, the drag force is proportional to the velocity ($v$), viscosity ($\eta$) and diameter of the ball ($d$) as

$$f = 3\pi \eta d v \qquad (1)$$

As the sphere is falling through the liquid, two forces are acting on the sphere. There is a buoyant force that is pushing upwards and a force due to gravity that is pulling the sphere downward. The buoyant force ($F_b$) is proportional to the volume of the fluid that is displaced by the submerged object ($\frac{4\pi}{3}r^3$ for a sphere of radius $r$), and the difference in density of the ball or object ($\rho_b$) and the fluid ($\rho_l$) as

$$F_b = \frac{4}{3}\pi r^3 (\rho_b - \rho_l) g \qquad (2)$$

Here $g$ is the acceleration due to gravity. The net driving force $F$ is then given by,

$$F = mg - F_b = \frac{4}{3}\pi r^3 (\rho_b - \rho_l) g \qquad (3)$$

When the drag force equals this net driving force ($F$), the ball will stop accelerating and will reach terminal velocity ($v_t$). Using the above three equations, we calculate the viscosity of the fluid, $\eta$, by making the assumption the ball has reached terminal velocity for a laminar fluid situation. The viscosity is given by

$$\eta = \frac{d^2}{18 v_t}(\rho_b - \rho_l)g \qquad (4)$$

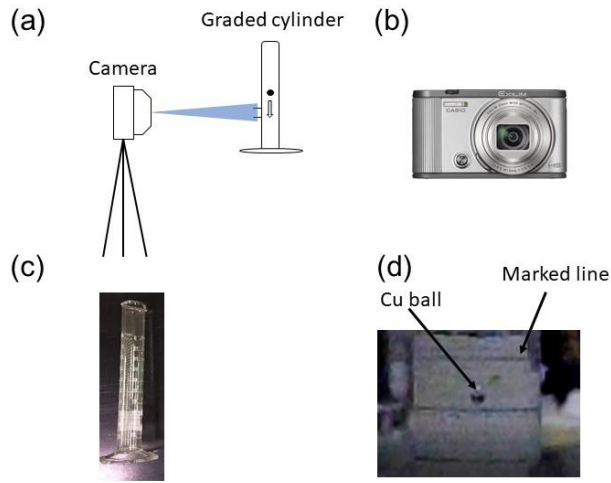

Figure 1: (a) Schematically drawn experimental setup. The video of the dropping ball was recorded by a high speed camera. The ball was dropped by a tweezer inside a graduated cylinder. (b) The camera (CASIO- EXILIM EX-ZR1700SR) used in this experiment. (c) The image of the graduated cylinder. A white paper with marked line was attached in the other side of the cylinder to facilitate the measurement. The white background makes easier to see the ball clearly and measure the time the ball takes to travel between marked lines. (d) One frame of the recorded video. The image is showing the image of the Cu ball and marked black lines drawn in a white paper that was attached to the cylinder.

**Experimental Section**

This experiment demonstrates a simple idea to measure the viscosity with high precision of a fluid by taking advantage of a commercially available high frame rate camera. We utilized common lab equipment such as a graduated cylinder, caliper, flask, tweezer, and scale. We estimated that our device cost ~$345 as shown in Table 1. The metallic Cu balls of different diameters were obtained from McMaster Carr.

We expect that the experiment is easy enough to be performed by an upper-level undergraduate student. The experimental setup is shown schematically in Fig.1a. The optical image of our high speed camera is shown in Fig.1b. First, we prepare a graduated cylinder filled with a test fluid (Fig.1c). To record the video of the ball drop, we mounted a high-speed camera on a tripod (Fig.1b) focused on the lower part of the graduated cylinder. Since we need to measure only the terminal velocity, we record the ball drop motion at the lower part of the cylinder. One video frame image of the motion of the ball captured is shown in Fig.1d. To facilitate the recording of the ball drop, we attached a parallel dark lines (1 cm separation) on a white paper attached to the other side of the cylinder as shown in Fig.1d. We calculate the velocity by measuring the time (or the number of frames) the ball takes to travel between two lines. The video frame was analyzed by Microsoft™ Window Media Player.

| Equipment | prices |
|---|---|
| 50 pieces of 0.8mm Cu ball | $12.98 |
| Caliper | $17.85 |
| 100 ml Graduated Cylinder | $6.99 |
| 500 ml beaker | $13.65 |
| 100 ml beaker | $2.19 |
| Precision digital scale | $60.00 |
| Tweezers | $0.91 |
| 1000 fps camera | $230.50 |
| total | $345.07 |

Table 1: The price of different elements of the experimental setup. Note that there are other models of CASIO-EXILIM family that supports also 1000 fps recording.

To tune the viscosity of a liquid, we used different concentrations of the glycerol-water mixture.[8,9] First, we determined the density of this mixture, $\rho_l$, by measuring the mass of the liquid. The volume of the mixture $V_{\text{mixture}}$ and the density ($\rho_l$) are given by,

$$V_{\text{mixture}} = \frac{m_{\text{glycerol}}}{\rho_{\text{glycerol}}} + \frac{m_{h_2 0}}{\rho_{h_2 0}} \tag{5}$$

$$\rho_l = \frac{m_{h_2 0} + m_{\text{glycerol}}}{V_{\text{mixture}}} = \frac{m_{\text{total}}}{V_{\text{mixture}}} \tag{6}$$

The glycerol percentage $P(\%)$ by mass of the mixture is given by.

$$P(\%) = \frac{m_{\text{glycerol}}}{m_{\text{glycerol}} + m_{\text{h}_2\text{0}}} \tag{7}$$

After measuring the diameter, *d*, of the Cu ball using a caliper, we pick the Cu ball with a tweezer and drop it in the center of the graduated cylinder containing a known percentage of water/glycerol mixture liquid. We measured the temperature of the fluid before dropping the ball.

The camera was set on recording this event and had a perpendicular view of the graduated cylinder. The camera settings were adjusted to the highest frame rate. We found that it is ideal to shoot in low light levels for clearer/sharper recording to eliminate the effect of in-line frequency (60 Hz) on-off of the room light bulbs.

The time was determined by counting the number of frames. The time for each frame is ~1/frame rate. The velocity of the ball was determined by measuring the number of frames the ball took to travel between two parallel dark lines (distance *L*).

**Results and Analysis**

The experiment consisting of glycerol percentages ranging from 50% to 100% glycerol solution. To study the effect of the ball size, we study three different sizes of Cu balls. We first measured the average velocity of the various balls all containing the same density shown in Fig.2. The velocity measurement was conducted six times to determine the average value. We found that the larger ball consistently has a larger velocity in lower glycerol percentages. As we increase the concentrations of glycerol, the measured velocity converges as shown in Fig.2.

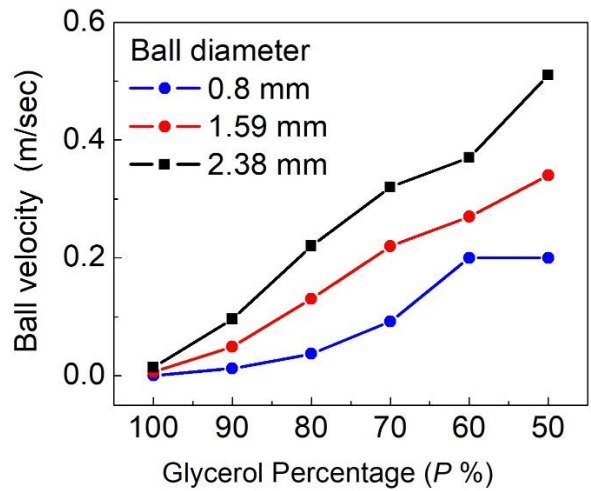

Figure 2: Measured velocity of a ball inside different percentage glycerol-water mixer. The figure is presenting the velocity data for three different sizes of the Cu balls.

After measuring the velocity, we can easily calculate the viscosity of the fluid using Eq.(4). We measured the viscosity for three different diameters of Cu balls. The results are shown in Fig.3. We also compare our measured value with reported viscosity values for different glycerol-water mixer by Segur and Obserstar.[10] The authors determined the viscosity value by using Cannon and Feske viscometer. We see that our measured value is in excellent agreement with the reported values as shown in Fig.3. We note that there is a deviation as the water concentrations are becoming larger. We speculate that this happens because the Cu ~~sphere~~ ball may not reach the terminal velocity for low-density liquid in our setup. This is also evident as the deviation from the reported values becomes larger for heavier, i.e. larger diameter, Cu ball. This deviation or error can be eliminated if we use a longer graduate cylinder or lighter weight Cu ~~sphere~~ ball or low-density metal sphere, e.g., sphere made of aluminum.

To measure the viscosity with high precision, it is important to utilize the proper ball size as low viscous fluids require a smaller diameter ball or lower density metal spheres. The smaller diameter ball can reach the terminal velocity inside the graduate cylinder used in this experiment. Another way, one can obtain accurate viscosity by using an aluminum sphere as it has a lower density than Cu.

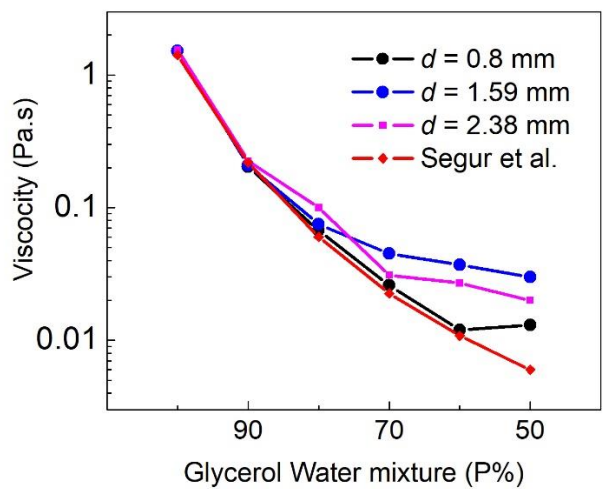

Figure 3: Measured viscosity values for different glycerol-water mixture. The measurement was conducted for three different sized Cu balls. The measured value is in good agreement with the value measured by other technique (see text for details).

**Conclusion**

Overall, we used an inexpensive method to measure the viscosity of a Newtonian fluid. The measurement technique described in this report is conveniently low-tech, but practically useful. The experiment described can be set up in a simple laboratory, including an undergraduate laboratory or high school teaching laboratory. The limitations of the ball drop method occur when the viscosity is too low so that the time range for the ball drop becomes significantly shorter than the frame time of the camera. But this limitation can be overcome by using a lower density metallic sphere or increasing the graduated cylinder length. The low-cost viscosity measurement technique presented in this paper is simple and reliable for practical applications.

**Acknowledgments**

A number of undergraduate students and visiting high school students participated in the experiment. A.K.M.N. acknowledges the support from the National Science Foundation Grant (ECCS-1708907) and Department of Defense Award (ID: 72495RTREP).